\documentclass[%
 aip,
rsi,%
 amsmath,amssymb,
preprint,
]{revtex4-1}

\usepackage{graphicx}% Include figure files
\usepackage{dcolumn}% Align table columns on decimal point
\usepackage{bm}% bold math
\begin{document}

%\preprint{AIP/123-QED}

\title[]{Time- and angle-resolved photoemission spectroscopy of solids in the extreme ultraviolet at 500 kHz repetition rate}% Force line breaks with \\
\author{M.~Puppin}
\email{michele.puppin@epfl.ch}
\affiliation{ 
Department of Physical Chemistry, Fritz Haber Institute of the Max Planck Society, Faradayweg 4-6, 14195 Berlin
}%
\affiliation{ Laboratory of Ultrafast Spectroscopy, ISIC, and Lausanne Centre for Ultrafast Science
(LACUS), \`Ecole Polytechnique F\`ed\`erale de Lausanne, CH-1015 Lausanne, Switzerland
}%

\author{Y.~Deng}
\affiliation{ 
Department of Physical Chemistry, Fritz Haber Institute of the Max Planck Society, Faradayweg 4-6, 14195 Berlin
}%
\affiliation{Paul Scherrer Institute, SwissFEL, 5232 Villigen PSI, Switzerland
}%

\author{C.~W.~Nicholson}
\affiliation{ 
Department of Physical Chemistry, Fritz Haber Institute of the Max Planck Society, Faradayweg 4-6, 14195 Berlin
}%
\affiliation{D\`epartment  de Physique and Fribourg Center for Nanomaterials, Chemin du Mus\`ee 3, Universit\`e de Fribourg, CH-1700 Fribourg, Switzerland
}%

\author{J.~Feldl}
\affiliation{ 
Department of Physical Chemistry, Fritz Haber Institute of the Max Planck Society, Faradayweg 4-6, 14195 Berlin
}%

\author{N.~B.~M.~Schr{\"o}ter}
\affiliation{ 
Department of Physical Chemistry, Fritz Haber Institute of the Max Planck Society, Faradayweg 4-6, 14195 Berlin
}%
\affiliation{ 
Paul Scherrer Institute, WSLA/202, 5232 Villigen PSI, 
Switzerland
}%

\author{H.~Vita}
\affiliation{ 
Department of Physical Chemistry, Fritz Haber Institute of the Max Planck Society, Faradayweg 4-6, 14195 Berlin
}%

\author{P.~S.~Kirchmann}
%\affiliation{ Department of Physical Chemistry, Fritz Haber Institute of the Max Planck Society, Faradayweg 4-6, 14195 Berlin }%
\affiliation{Stanford Institute for Materials and Energy Sciences, SLAC National Accelerator Laboratory, 2575 Sand Hill Road, Menlo Park, California 94025, USA
}%

\author{C.~Monney}
\affiliation{ 
Department of Physical Chemistry, Fritz Haber Institute of the Max Planck Society, Faradayweg 4-6, 14195 Berlin
}
\affiliation{D\`epartment  de Physique and Fribourg Center for Nanomaterials, Chemin du Mus\`ee 3, Universit\`e de Fribourg, CH-1700 Fribourg, Switzerland
}%
\author{L.~Rettig}
\affiliation{ 
Department of Physical Chemistry, Fritz Haber Institute of the Max Planck Society, Faradayweg 4-6, 14195 Berlin
}%

\author{M.~Wolf}
\affiliation{ 
Department of Physical Chemistry, Fritz Haber Institute of the Max Planck Society, Faradayweg 4-6, 14195 Berlin
}%

\author{R.~Ernstorfer}
\email{ernstorfer@fhi-berlin.mpg.de}
\affiliation{ 
Department of Physical Chemistry, Fritz Haber Institute of the Max Planck Society, Faradayweg 4-6, 14195 Berlin
}%

%\date{\today}% It is always \today, today,
             %  but any date may be explicitly specified

\begin{abstract}
Time- and angle-resolved photoemission spectroscopy (trARPES) employing a 500~kHz extreme-ultraviolet (XUV) light source operating at 21.7~eV probe photon energy is reported. Based on a high-power ytterbium laser, optical parametric chirped pulse amplification (OPCPA), and ultraviolet-driven high-harmonic generation, the light source produces an isolated high-harmonic with 110~meV bandwidth and a flux of more than $10^{11}$ photons/second on the sample. Combined with a state-of-the-art ARPES chamber, this table-top experiment allows high-repetition rate pump-probe experiments of electron dynamics in occupied and normally unoccupied (excited) states in the entire Brillouin zone and with a temporal system response function below 40~fs.  
\end{abstract}

\maketitle

\section{\label{sec:Introduction}Introduction}

The temporal evolution of the microscopic properties of a solid brought out of equilibrium by an ultrashort laser pulse provides fundamental insights into the couplings between its electronic, spin and lattice degrees of freedom. Time-resolved spectroscopy allows disentangling experimentally the interplay and coupled evolution of these subsystems, whose characteristic timescales are set by their microscopic interactions. Ultrashort light pulses can resolve the fast electronic evolution occurring on femtosecond timescales. Time- and angle-resolved photoemission (trARPES) directly accesses electronic states of a material with momentum resolution, as the system is driven out of equilibrium by a femtosecond optical pump pulse. ARPES measures the angular distribution and the kinetic energy of photoemitted electrons: it is frequently assumed that the ARPES intensity I(\textbf{k},$\omega$) can be written as the product between the occupation probability of the electronic state f(\textbf{k},$\omega$), the single-particle spectral function A(\textbf{k},$\omega$) and a matrix element between the initial and final state $\lvert M_{if}^{\textbf{k}} \rvert ^2$; here $\textbf{k}$ and $\omega$ denote the electron's wavevector and angular frequency, respectively. Many-body effects are encoded in the spectral function A(\textbf{k},$\omega$) and manifest themselves in renormalization of the bare electronic bands and in the observed lineshape\cite{Andrea2004}.
In a trARPES experiment, the distribution I(\textbf{k},$\omega$) is collected for a series of delays ($\tau$) between pump and probe pulses: after perturbation, the population distribution f(\textbf{k},$\omega$,$\tau$) evolves towards a quasi-thermal distribution and energetically relaxes on femto- to picosecond timescales\cite{Lisowski2004}. During relaxation, the concomitant many-body interactions affect the transient spectral function A(\textbf{k},$\omega$,$\tau$) and even the photoemission matrix elements might change, if the final state's orbital symmetry is altered\cite{Boschini2018}. trARPES accesses at once the population dynamics, the evolution of the spectral function and the evolution of matrix elements. trARPES has found increasingly successful applications in the past few decades\cite{HAIGHT1995review,Bovensiepen2012,Smallwood2016a}: among many examples, trARPES was used to study photo-induced phase transitions\cite{Schmitt2008,Rohwer2011,Monney2016,zong_evidence_2018,ChrisCr2016} and to observe electronic states above the Fermi level, unoccupied under equilibrium conditions\cite{bertoni2016,hein2016,Wallauer2016,Sobota2012a,Mahmood2016}. Energy conservation in the photoemission processes imposes that a femtosecond light source for trARPES must possess a photon energy $\hbar \omega_{ph}$ exceeding the work function $\Phi$, which in most materials lies in the range between 4 to 6~eV. Ultraviolet femtosecond light sources are thus required for these experiments. The conservation of the electrons' in-plane momentum ($\hbar k_{\parallel}$) in the photoemission process allows reciprocal space resolution. The advantage of a probe with high photon energy is the increased range of observable reciprocal space: low-photon-energy sources are limited to regions close to the Brillouin zone center.

The scope of this work is to describe an experimental setup for trARPES based on a newly developed light source which operates at an energy of 21.7~eV and provides a monochromatized photon flux exceeding $10^{11}$ ph/s on the sample. 
The light source is embedded in a beamline equipped with a state-of-the-art ARPES end-station, where trARPES experiment can be performed with a system response function (pump-probe cross-correlation) better than 40~fs, a source linewidth of 110~meV and at a repetition rate of 500~kHz. This is achieved by performing a change of the employed laser technology, from a conventional titanium:sapphire laser to an optical parametric chirped pulse amplifier (OPCPA), entirely based on sub-picosecond ytterbium lasers\cite{Puppin2015}. The experimental apparatus presented in this work bridges the existing technology gap between widespread high-flux, high repetition rate sources with low photon energy  \cite{Smallwood2012_corr,Ishida2014,Koralek2007,Boschini2014b,Sobota2012a,Faure2012_corrected} and conventional high photon energy sources, based on high-order harmonic generation and operating at lower repetition rates \cite{Haight1994a,Mathias2007,Dakovski2010_corr,Turcu2010,Frietsch2013_corr,Eich2014a,Plogmaker2015_corr,Rohde2016,Ojeda2016}, thereby enabling a vast class of new experiments to be performed in the whole Brillouin zone of most materials. 
The structure of the paper is as follows: section \ref{sec:light source} will be dedicated to the description of the light source, whereas the trARPES beamline will be described and characterized in section \ref{sec:beamline}.

\section{\label{sec:light source}The light source}
%\subsection{\label{sec:reprate}High-repetition rate light source for trARPES}
For each trARPES experiment, a multi-dimensional data set is recorded: the photoelectron intensity distribution is measured as a function of energy, parallel momentum and pump-probe delay time. In order to collect sufficient statistics, data have to be accumulated over numerous laser pulses since the probe pulse intensity cannot be increased arbitrarily. In fact, multiple photoelectrons per pulse lead to space-charge effects \cite{Passlack2006,hellmann2009,Graf2010}: the photoemitted electrons are initially confined in a small volume leading to strong Coulomb repulsion and a spreading of the photoelectron cloud, worsening the energy and momentum resolution. The best way to mitigate this effect, is to reduce the photon flux and accordingly increase the experimental repetition rate to compensate for the reduced count rate. 

However, in pump-probe experiments, the pump excitation also has to be taken into consideration when designing the experiment. First, it is important that the sample re-equilibrates within the laser's duty cycle; second, in many experiments a minimum energy density has to be reached to initiate a certain non-equilibrium process, for example to photo-induce a phase transition \cite{Schmitt2008,Monney2016,Nicholson2018}. In many cases, the photo-degradation of the sample occurs faster at higher repetition rates, making a vast variety of studies unfeasible. The ideal repetition rate depends therefore on the sample under investigation, however, the need for high photon energies, complicates trARPES when going beyond a few tens of kHz.
This stems from the fact that femtosecond lasers, typically in the visible to near-infrared (VIS-NIR) spectral ranges, have to undergo a high-order frequency up-conversion process, requiring laser amplifiers with high peak powers. Titanium:sapphire laser amplifiers are currently the workhorse of femtosecond science, providing terawatt pulses at the kHz level\cite{Backus1998a}. Intensities on the order of $10^{14}$~W/$cm^2$ are easily reached, routinely  enabling the generation of extreme ultraviolet radiation (XUV, 20-100~eV) via high harmonic generation (HHG)\cite{Popmintchev2010}. 

Unfortunately, it is difficult to scale XUV generation with Ti:sapphire lasers above a repetition rate of a few tens of kHz: some of the highest repetition rates demonstrated so far for trARPES experiments are 50-100~kHz\cite{Buss2018,Wallauer2016}, whereas static ARPES experiments have been demonstrated at the MHz level\cite{Chiang2015b}. Several approaches have been devised to increase the repetition rate of table-top light sources, often employing different type of femtosecond lasers: for example, cavity-enhanced XUV generation, provide repetition rates of several tens of MHz and space-charge-free photoemission\cite{mills2015,corder2018,ishida2016}; in this regime also the pump excitation has to be modest to allow for sample's relaxation. In the present work, we aim at increasing the repetition rate of single-pass harmonic generation by employing an OPCPA based on ytterbium lasers, with a sufficient average power to efficiently drive the process at 500 kHZ. 

\subsection{\label{sec:OPCPA}Optical parametric chirped pulse amplification}

In the past decades, several ytterbium lasers operating at 100s of Watts of average power and MHz-level repetition rates were demonstrated\cite{Russbueldt:09}: such lasers are already capable of directly producing XUV radiation via high-harmonic generation \cite{Boullet2009}, or can be brought to the few-cycle pulse regime by nonlinear compression \cite{Hadrich2014b}. Short pulse duration and high mode quality make amplified ytterbium lasers ideally suited for OPCPA\cite{Vaupel2013c}. In OPCPAs, by controlling the spectral phase of the amplified optical pulses, the central frequency and the bandwidth can be tuned at will. In the VIS-NIR range, several ytterbium-based OPCPAs were demonstrated either with a broad bandwidth spectrum, supporting few cycle pulses\cite{Andersen2006a,Homann2008,Emons2010a,Schulz2011,Nillon2014} or with a frequency-tunable spectrum over a wide range\cite{Schriever2008,Pergament2014,Riedel2014,Bradler2014,Puppin2015}. 
This combination of high average power and time-bandwidth flexibility, make OPCPAs promising candidates for the next generation of femtosecond lasers\cite{Fattahi2014a} and high-repetition rate sources\cite{Krebs2013}.

\begin{figure*}
	\includegraphics{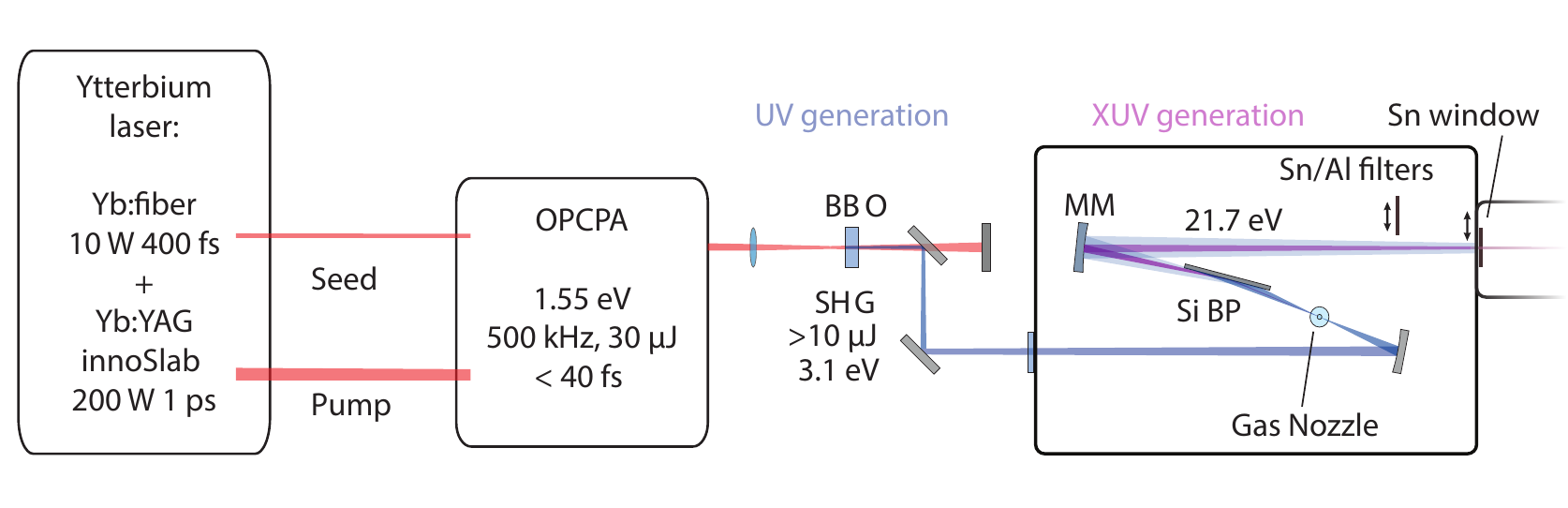}
	\caption{\label{fig:hhgsummary} Schematic layout of the 21.7~eV light source. OPCPA: optical parametric chirped pulse amplifier, SHG: second harmonic generation, BP: Brewster Plate, MM: Multilayer mirror.}
\end{figure*}

The approach followed in this work comprises an OPCPA based on an Yb:YAG laser which produces femtosecond NIR light at 1.55~eV (figure \ref{fig:hhgsummary}) at 500~kHz. The details of the laser system has been described elsewhere\cite{Puppin2015} and will only be summarized here. The OPCPA is seeded by a broad-bandwidth white light continuum, which is generated in a YAG crystal\cite{Bradler2009} by a fiber laser with a pulse duration of 400~fs (full-width at half maximum, FWHM) at a carrier wavelength of 1030~nm. 
The pump for the parametric amplification is the second harmonic (515~nm) of a slab amplifier \cite{Russbueldt:09} capable of 200~W average power and with picosecond pulse duration. This pump laser is seeded by the same fiber laser producing the broadband seed for the OPCPA: this ensures an inherent optical synchronization between the seed and the pump pulses, important for parametric amplification of sub-ps pulses \cite{riedelOE2013}.

A single amplification stage in a beta-barium borate ($\beta$-BBO) crystal is sufficient to saturate the pump conversion efficiency, simplifying the optical setup. The central frequency of the amplifier, tunable within 650~nm - 950~nm, was set to 800~nm for practical reasons, to access the widespread optical components available for Ti:sapphire lasers.
After a prism compressor, 30~$\mu$J pulses at 500~kHz are available for further conversion to the XUV range. For typical trARPES experiments, the spectrum of the laser is set to a bandwidth of 80~meV (FWHM), around the photon energy of 1.55~eV and compressed to a pulse duration below $35$~fs (FWHM) with a mode quality $M^{2}<1.5$. 

\subsection{\label{sec:level1}UV-driven high-harmonic generation at 500 kHz}

In the process of high-harmonic generation in gases, the XUV radiation originates from electrons recombining with their parent ion after being accelerated in the optical field\cite{chang2011fundamentals}. 
The hall-mark of this effect in the frequency domain is the appearance of odd harmonics of the fundamental laser frequency $\omega_{0}$, co-propagating with the driving radiation. The irradiance for several harmonic orders is nearly constant in a wide energy region, the so-called plateau region, which extends up to a cut-off energy determined by the atomic ionization potential and the electron's ponderomotive potential \cite{Lewenstein1994b}. The temporal structure of the harmonics, consisting of bursts of radiation at every half-cycle of the driving electric field, has proven to be extremely valuable for femtosecond\cite{HAIGHT1995review,Bovensiepen2012,Smallwood2016a} and attosecond spectroscopies\cite{krauz2009atto}.

Although it is possible to directly generate high harmonics with the NIR pulses of the OPCPA, there are advantages in frequency up-converting the pulses in a nonlinear crystal and generating high harmonics with UV pulses\cite{Rohde2016,Wang2015a,Eich2014a}.
The conversion losses introduced are largely compensated by an increase of the single atom response\cite{Lai2013} and an improvement of macroscopic phase-matching conditions\cite{Wang2015a}. At a given XUV energy, the doubled driver's photon energy reduces the order $q$ of the nonlinear processes. This leads to a lower Gouy phase for the focused laser beam and a lower phase mismatch resulting from the free-electron plasma: an overall efficiency increase of nearly two orders of magnitude was reported using Ti:sapphire lasers\cite{Wang2015a}. On top of this increased efficiency, UV-driven HHG simplifies selection of a single harmonic out of the XUV frequency comb, as will be  be explained in more detail in section \ref{sec:mono}.

\subsection{High-pressure gas target for HHG in a tight-focusing geometry}

A high photon flux for trARPES can only be achieved if the harmonic radiation produced across a macroscopic volume of the gas target adds coherently\cite{Popmintchev2010}. 
This phase-matching condition is more difficult to achieve in a tight-focusing regime\cite{Heyl2012}: this is a direct consequence of the Gouy phase of the laser beam, which becomes an increasingly important term as the focal spot gets smaller. 
A detailed analysis of the scaling properties of the phase matching relations for HHG \cite{Rothhardt2014,Heyl2012,Heyl2016}, reveals that the theoretical conversion efficiency can be made invariant of the spot size $w_0$, provided that the pulse energy $\epsilon$, the gas pressure p and the gas target length $L$ are suitably rescaled. More precisely\cite{Rothhardt2014}, the efficiency reached in a loose focusing geometry at certain intensity can be also expected for a lower pulse energy $\epsilon'=\epsilon  s$ and a smaller spot size $\sqrt{s} w_0$, where $s<1$ is the scaling factor. For this to happen, a higher pressure $p'= p s^{-1}$ has to be achieved in a gas target confined in a shorter length $L'=L s$: the main challenge of a tight-focused HHG light source is therefore to realize a high-pressure, confined gas target within a high-vacuum beamline\cite{Comby2018}.
\begin{figure}
	\includegraphics{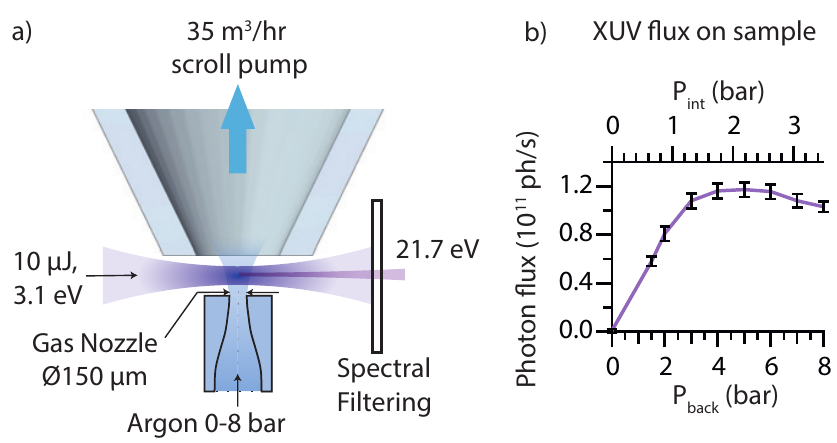}
	\caption{\label{fig:gasfigure} a) Side view of the differential pumping system for the high-pressure gas nozzle. b) Measured flux at 21.7~eV on the sample as a function of the backing pressure $P_{back}$, the estimated pressure at the focus of the laser beam $P_{int}$ is shown in the mirror axis.}
\end{figure}

In our setup this is realized by focusing the beam at the output of a small diameter converging nozzle\footnote{Biomedical instruments GmbH, glass nozzles of type Wieland, see \url{https://biomedical-instruments.com}}, connected to a gas line with a pressure of several bars: the gas nozzle configuration is schematized in figure \ref{fig:gasfigure}a). The gas jet freely expands in a high-vacuum chamber: to minimize absorption losses, the lowest residual pressure is desirable along the beam path of the strongly ionizing XUV radiation. To reduce the overall gas load, a skimmer with an aperture of 3~mm diameter is placed in front of the nozzle. A three-axis manipulator holds the skimmer and is used for precise alignment. A scroll pump with a pumping speed of 35~$m^3$/hr is connected to the rear-side of the skimmer through a flexible hose. For a typical nozzle diameter of 150~$\mu$m, the differential pumping setup ensures long-term operation of the 700~l/min turbomolecular pump of the vacuum chamber, with up to 9~bar backing pressure of argon; the chamber's pressure during operation is in the $10^{-2}$ mbar range. 
The beam path following the gas target consists of several optical components to re-collimate the beam, suppress the 3.1~eV UV driver and isolate a single harmonic around 22~eV.

\subsection{\label{sec:mono}Single harmonic selection}

The separation between neighboring harmonics is $2 \omega_{0}$: in UV-driven HHG, the spacing between different orders also increases, facilitating the monochromatization of the XUV radiation\cite{Wang2015a,Rohde2016}. For a trARPES experiment, a single harmonic must be isolated with good spectral contrast: the contamination of neighboring harmonics produces unwanted replicas of the photoelectron spectrum. In particular, if the $q$-th harmonic is selected, a residual $q+2$ order produces photolectrons from deeper valence states which overlaps in energy with states in the vicinity of the Fermi level. 

Monochromators based on a single diffraction grating, widely adopted in XUV synchrotron beamlines, have the drawback of introducing a pulse front tilt which hinders the temporal resolution in pump-probe experiments. Special designs must be adopted to minimize this effect without sacrificing the transmission considerably \cite{Frassetto2011,Frietsch2013_corr}. The alternative approach is to use a combination of reflective and transmissive optics to isolate a single harmonic from the fundamental beam.

\begin{figure}
	\includegraphics{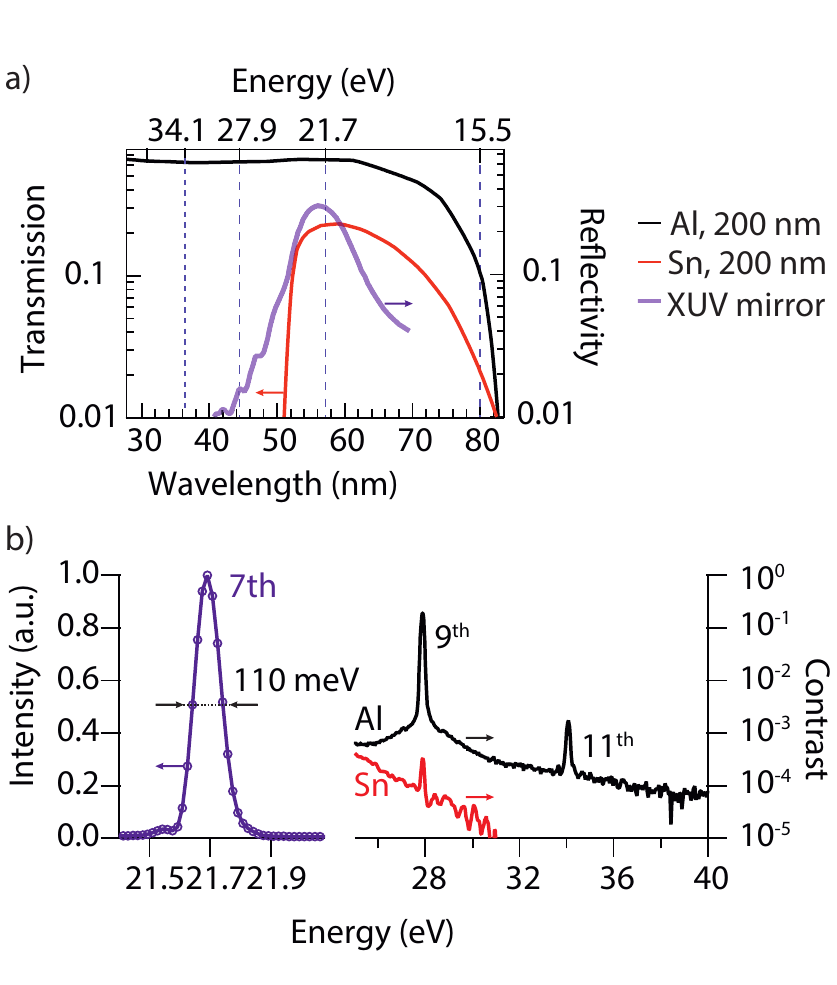}
	\caption{\label{fig:monochromator} a) Left vertical axis: Theoretical transmission\cite{HENKE1993} of 200-nm-thick Sn (red line) and Al (black line) foils. Right vertical axis: theoretical reflectivity of the multilayer mirror. The vertical dashed lines indicate the position of the harmonics, from order 5 to 11. b) Left vertical axis: Normalized spectrum of the $7^{th}$ harmonic at 21.7~eV, highlighting the FWHM of 110~meV. Right vertical axis: intensity of the $9^{th}$ and $11^{th}$ harmonics, relative to the main 21.7~eV line. The contrast of Al (black line) and Sn (red line) are compared.}
\end{figure}

In this kind of monochromator, a harmonic is selected by multilayer-coated mirrors, designed for the specific wavelength: this preserves the pulse duration, sacrificing however the tunability. Multilayer mirrors have a typical reflectivity of few tens percent at the wavelength for which they are designed; unfortunately, away from the reflectivity peak, a non-negligible residual reflectivity  decreases the contrast between neighboring harmonics. In the case of NIR-driven HHG the contrast is improved by a second, lossy reflection \cite{Mathias2007}. A transmissive filter is required to fully suppress the co-propagating fundamental radiation, which is several orders of magnitude brighter than the harmonics. A free-standing metal foil with a thickness of some hundreds of nanometers effectively suppresses the fundamental radiation and still transmits a reasonable proportion of the XUV radiation. Aluminum is commonly used, as it acts as a high-pass filter above 20~eV: the estimated \cite{HENKE1993} transmission of 200~nm Al is shown in figure \ref{fig:monochromator}a). 

The spacing between UV-driven harmonics (6.2~eV in our case) allows for a simplified setup based on a single reflection on a multilayer mirror, followed by a transmission through a free-standing tin foil \cite{Wang2015a}. Tin has a transmission window centered approximately at 22~eV, close to the $7^{th}$ harmonic of the 3.1~eV driver: the theoretical transmission through 200~nm of Sn is shown in figure \ref{fig:monochromator}a) together with the position of the closest harmonics (vertical dashed lines).
Experimentally it was determined that a Sn foil (Lebow) of 200~nm nominal thickness had a transmission of 9\% at a photon energy of 21.7~eV. The reduced transmission compared to the theoretical value is likely due to oxide layers forming at the surfaces. 

The optical setup following the gas target is shown schematically in figure \ref{fig:hhgsummary}. High-repetition rate systems have an inherently higher thermal load which could easily damage the thin metal filter. To solve this problem, a Brewster plate consisting of a silicon single crystal was used as first optical element in the beamline, placed approximately 100~mm after the focus. Most of the fundamental power is absorbed in the silicon wafer, preventing damage in the following components. The thermal load did not cause a sensible mode degradation of the XUV beam: the characterization of the XUV beam profile is discussed in section \ref{sec:beamline}. The reflectivity of the Si plate in the XUV was estimated numerically using the free software IMD\cite{Windt1998}, including the effects of a passivating SiO$_2$ layer of about 1~nm, as expected for a commercial high-purity silicon single crystal\cite{doering2008handbook}. At the Brewster angle for 400 nm (80$^{\circ}$ angle of incidence) the measured reflected power for the driving radiation is below 100~mW, while the calculated reflectivity for the 7th harmonic is 80\%. 

After the silicon wafer, the XUV radiation is re-collimated by a spherical mirror with a focal length of 200~mm, coated with a multilayer with reflectivity centered at the $7^{th}$ harmonic\footnote{Reflective X-ray Optics LLC, see \url{http://www.rxollc.com/}}. The theoretical reflectivity of the XUV mirror is on the order of 30\% and is plotted in in figure \ref{fig:monochromator}a). The mirror's coating consists of a silicon layer, covering a multilayer composed of chromium and scandium, realized on an XUV-grade substrate (flatness $\lambda/20$, roughness $<0.2$~nm RMS). 
The metallic Sn filter is mounted as a window of a gate valve and a motorized filter wheel can be used to insert additional filters (Al or Sn, 200 nm thickness).

A gold mirror can be inserted in the beam to reflect the beam into a grating spectrometer\footnote{McPherson 234/30, equipped with a CCD camera, ANDOR IDUS DO420A-OE-995 and a concave platinum-coated grating with 2400~lines/mm} to measure the XUV spectrum. A typical spectrum is shown in figure \ref{fig:monochromator}b). The lower limit for the instrumental resolution under the measurement conditions is 60~meV: the measured line width of the $7^{th}$ harmonic is 110~meV. The contrast between the $7^{th}$ harmonic at 21.7~eV and the $9^{th}$ harmonic at 27.9~eV was measured with a single 200~nm-thick Al filter, with a flat transmission above 22~eV and compared with the one of a single 200~nm-thick Sn foil. The contrast between the $7^{th}$ harmonic and the $9^{th}$ improves by more than two orders of magnitude in the latter case: an additional 200~nm-thick Sn foil reduces the signal at 28~eV to the noise level of the spectrometer's detector, corresponding to a flux on the sample below $10^{7}$~ph/s. 

To characterize the $7^{th}$ harmonic radiant power, an XUV diode\footnote{AXUV100Al, Opto diode corporation, read by a picoammeter, Keithley 6485} mounted on a linear transfer arm can be inserted in the beam. It is important to note that no additional optical elements are present in the beamline after the XUV diode (see section \ref{sec:beamline}): the radiant flux measured in this position corresponds therefore to the one at the sample's position during trARPES experiments. The measured photon flux at 21.7~eV, calculated using the diode's factory responsivity of 0.15~A/W, is plotted as a function of the backing pressure $P_{back}$ of the nozzle in figure \ref{fig:gasfigure}b). Optimal phase matching for a 150~$\mu$m-diameter nozzle was observed at a backing pressure of approximately 5~bar for Argon. The gas pressure in the interaction region $P_{back}$ is shown in the mirror axis of figure \ref{fig:gasfigure}: this results assumes an ideal supersonic expansion of the gas\cite{pauly2000atom}, the minimal distance of the beam axis from the nozzle front face is assumed to be twice the beam radius $w(z)$, with a beam waist of $2w_0=25$ $\mu$m. The overall transmission of the transmissive monochromator with a single Sn foil is estimated to be 2.2\%: this allows to calculate the photon flux from the source before monochromatization, which is on the order of $5.6\times10^{12}$ photons/second for the $7^{th}$ harmonic.

\subsection{\label{sec:results}Comparison between different nozzles and noble gases}
The optimal gas medium length was studied by testing nozzles of different throat diameter (40, 80 , 150 and 500 $\mu$m), under the same focusing conditions in argon. The position of each nozzle was optimized to maximize the radiant flux with a pressure up to 9~bar, which was the maximal pressure for the gas fittings in the gas line. The best results for each configuration are reported in table \ref{table:hhg_nozzlecomparison}. A saturation of the XUV flux with pressure was observed only for the 150~$\mu$m-diameter nozzle, whereas in the case of the 500~$\mu$m-diameter nozzle, the turbomolecular pump overloaded before HHG saturation. For the 40~$\mu$m-diameter nozzle the signal was still monotonically increasing at the maximum backing pressure. 
The highest flux was measured with the smallest nozzle, indicating that for a medium length of $\approx$100~$\mu$m the harmonic re-absorption already limits the signal build-up. 
In practice and in view of performing long measurements, a 150~$\mu$m-diameter nozzle was preferred as HHG was found to be less sensitive to beam-pointing drifts. 
With this nozzle size and a typical backing pressure of 2~bar, the argon gas flow is below 300~sccm/min.

\begin{table}[]
	\centering
	\caption{Relative maximum 21.7~eV flux for different nozzle throat diameters, together with the measured flux on the sample. The source flux is calculated from the estimated monochromator transmission of 2.2\%.}
	\label{table:hhg_nozzlecomparison}
	\begin{tabular}{lllll}
		\hline \hline
		Nozzle throat ($\mu$m) &  40  & 80  & 150  & 500  \\ \hline 
		Relative flux &  1.00 &  0.74 & 0.60 & 0.13 \\ \hline
		Source flux ($\times 10^{12}$ ph/s) &  9.2 &  6.8 & 5.6 & 1.2 \\ \hline
		Flux on the sample ($\times 10^{11}$ ph/s) &  2.0 &  1.5 & 1.2 & 0.3 \\ \hline
	\end{tabular}
\end{table}

Other noble gases were also tested for the 150~$\mu$m nozzle: in each case, pressure and nozzle position were optimized to maximize the XUV flux. The relative intensities are listed in table \ref{table:hhg_gascomparison}. The best results were obtained with argon, even though krypton is expected to yield a stronger single-atom response\cite{Wang2015a,Wallauer2016}. It is possible that the higher ionization in the case of Kr prevented phase matching at the pulse peak, resulting in a shorter coherence length. A longer focal length could not be tested in the setup due to geometrical constrains, nonetheless, it is expected that a higher XUV radiant power would be possible, using the present driver laser and krypton as gas medium. Due to the scarcity and higher cost of this gas, a gas recycling system\cite{Wallauer2016,CHIANG201515} is needed for a high pressure nozzle, complicating its adoption in the current setup. 
Overall, the source's flux using Ar exceeds $10^{11}$~ph/s at 21.7~eV and at 500 kHz, which is well suited for trARPES experiments\cite{Nicholson2018}.
\begin{table}[tbp]
	\centering
	\caption{Comparison of the relative radiant flux for the 21.7~eV harmonic with different gases for 150~$\mu$m\ nozzle diameter and measured on the sample's position. The source's flux is calculated from the estimated monochromator transmission of 2.2\%.}
	\label{table:hhg_gascomparison}
	\begin{tabular}{llll}
		\hline \hline
		Noble gas &  Ne &  Ar & Kr \\ \hline 
		Relative flux &  0.02 &  1.0 & 0.78   \\ \hline
		Source flux ($\times 10^{12}$ ph/s) &  0.1 & 5.6 & 4.3 \\ \hline
		Flux on the sample ($\times 10^{11}$ ph/s) &  0.02 &  1.2 & 0.9 \\ \hline
	\end{tabular}
\end{table}

\section{\label{sec:beamline} The \lowercase{tr}ARPES beamline}
\begin{figure}
	\includegraphics{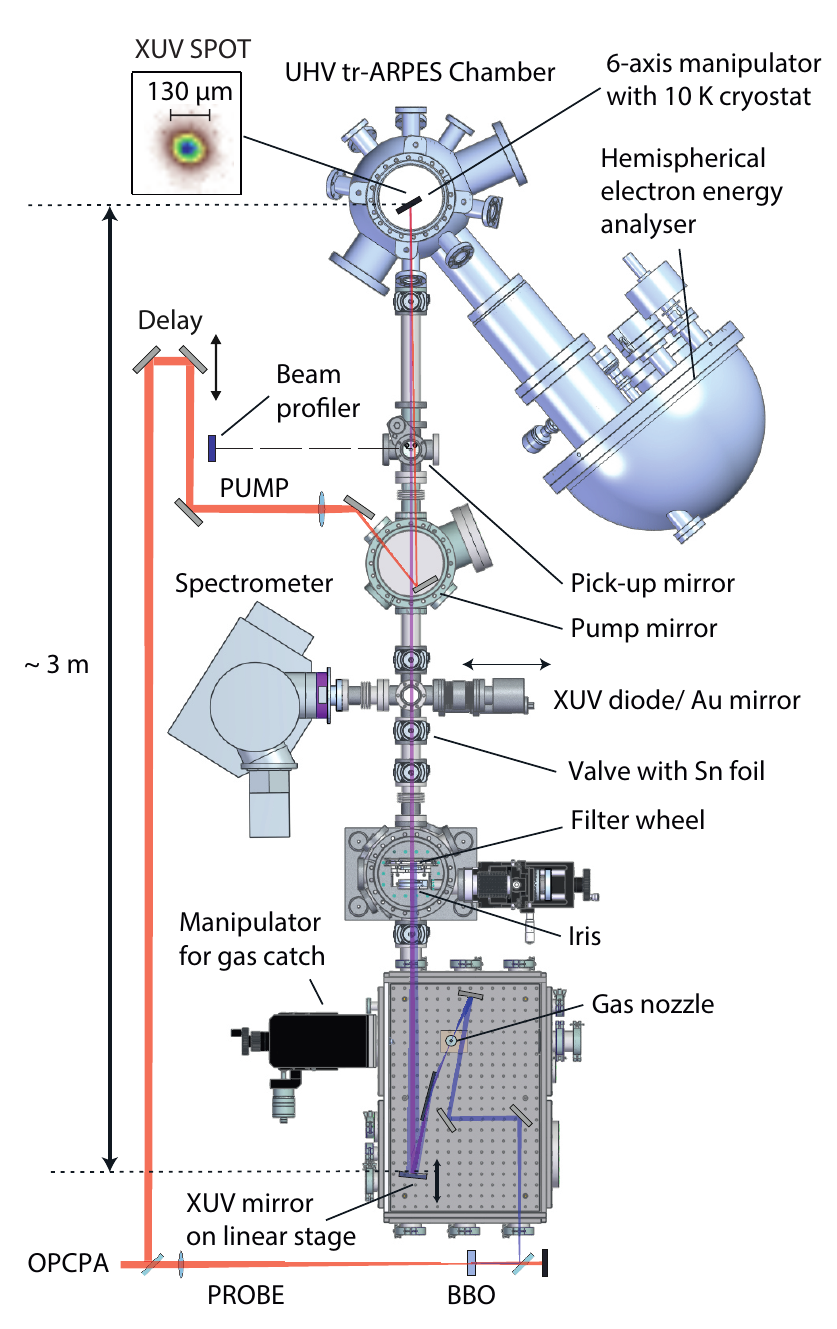}% Here is how to import EPS art
	\caption{\label{fig:beamline} Scaled layout of the beamline. Refer to Sec. \ref{sec:beamline} in the main text for a detailed description. Inset: Measured XUV profile.}
\end{figure}
An XUV beamline connects the light source to the experimental chamber: the setup is depicted in figure \ref{fig:beamline}. The XUV beam, generated in the high-pressure gas target in the HHG chamber, is reflected by the silicon Brewster plate and relayed by the XUV mirror towards a second chamber. In this chamber, a manual 3-axis manipulator holds a motorized iris used in the experiments to attenuate the XUV flux without changing the phase-matching conditions. This chamber also hosts a motorized filter wheel with 6 slots, containing Sn and Al filters, which can be inserted in the beam to improve the contrast between neighboring harmonics. After the filter wheel, a gate valve hosts the Sn foil, used both as transmissive filter for selecting the 7$^{th}$ harmonics and as a pressure barrier before the ultra-high-vacuum (UHV) experimental chamber. 

The next section of the beamline contains tools for HHG characterization: after the Sn foil a linear translation stage hosts the XUV diode, which measures the flux incident on the sample. The same linear translation arm holds a gold mirror, used to reflect the harmonics into the XUV spectrometer. A third vacuum chamber along the beamline hosts a fixed mirror mount, used to in-couple the pump beam, at an angle $\approx3^{\circ}$ relative to the XUV beam. 

The pump and the probe beams finally arrive to the UHV trARPES chamber where they overlap on the sample. The trARPES chamber reaches a base pressure in the upper $10^{-11}$~mbar range: thanks to the Sn window no significant increase is observed during measurements, when the high-pressure gas nozzle is operating. The XUV focal spot was characterized using a micro-channel-plate electron multiplier, followed by a phosphor screen imaged by a camera. The resulting beam profile is shown in the inset of figure \ref{fig:beamline} and measures 130~$\mu$m full-width at half maximum.  

The UHV chamber is equipped with a hemispherical electron energy analyzer\footnote{SPECS Surface Nano Analysis GmbH, product spectrometer PHOIBOS\textsuperscript{TM}
150 (2013), see \url{http://www.specs.de/cms/front_content.php?idart=122}}. The sample is installed on a 6-axis manipulator\footnote{SPECS Surface Nano Analysis GmbH, product manipulator CARVING\textsuperscript{TM}
(2013), see \url{http://www.specs.de}} which allows liquid helium cooling down to 10 K. The XUV beamline is connected to the analysis chamber in the same plane of the analyzer's entrance slit; the analyzer's input axis is fixed at an angle of $40^{\circ}$ relative to the beamline. The manipulator, holding the sample during the experiments, can be moved to an upper chamber, where samples are stored and prepared before the measurements. The samples are inserted in vacuum through a load-lock chamber and are moved to the sample storage chamber using a magnetic transfer arm.

Just before the trARPES chamber, a cross joint holds a linear translation stage with a metallic pick-up mirror: this can be inserted in the beam path to simultaneously reflect the pump and the residual 400~nm HHG driver, which is still observable if the Sn foil is removed. This allows for characterization of the pump beam profile and coarse temporal overlap of pump and probe beams on a photodiode. The XUV beam is aligned in the photoemission chamber by controlling a motorized mirror mount holding the multilayer mirror. The mount is translated along the beam axis with a linear translation stage, to precisely focus the source on the sample. Once the XUV beam is aligned relative to the electron energy analyzer focus, the pump beam can be overlapped on the probe beam by imaging with a CCD camera a Ce:YAG scintillator, installed on the manipulator. 

\section{\label{sec:TRA}Characterization of the \lowercase{tr}ARPES setup}
\begin{figure*}
	\includegraphics{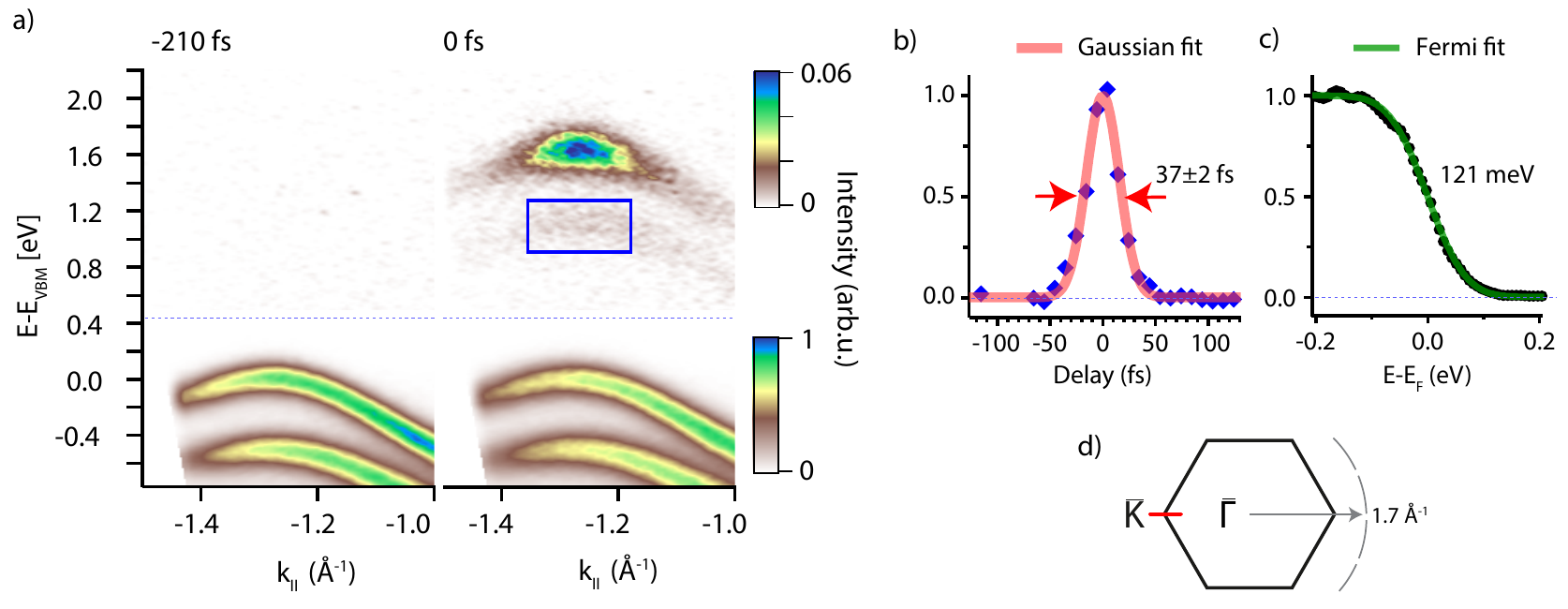}
	\caption{\label{fig:PESsummary} a) False color plots of trARPES data~210 fs before and during temporal overlap in WSe$_2$ at the $\overline{K}$ point. The photoelectron energy E is referenced to the top of the valence band at the $\overline{K}$ point, E$_{V\!B\!M}$. Above 0.45~eV, marked by a black line, a different color scale was used in the image to allow visualizing the weaker excited states signal, two color bars indicate the signal level, the same arbitrary units are used for both the images. The blue box indicates the integration area used to estimate the pump-probe cross-correlation. b) Integrated intensity of the photoemission signal in the blue box of panel a), showing the result of a fit using a Gaussian function, indicating a temporal cross-correlation between pump and probe signal of $37\pm2$~fs FWHM. c) Energy distribution curve at the Fermi edge of TbTe$_{3}$, at a temperature of 100 K. A Fermi function at a temperature of 100 K was convoluted with a Gaussian and fitted to the data, yielding a broadening of 121 meV. d) Surface-projected Brillouin zone of WSe$_2$. The red line indicates the position in reciprocal space of the ARPES plots of panel a). The gray line indicates the accessible momentum space.}
\end{figure*}

First trARPES experiments were demonstrated on WSe$_2$, a member of the well-studied transition-metal dichalcogenide semiconductors family \cite{Rettenberger1997,hein2016, Wallauer2016, bertoni2016}: the results are summarized in figure \ref{fig:PESsummary}. The photon energy of the light source is sufficient to reach electrons with a parallel momentum of 1.7 \AA$^{-1}$ (with a photoemission angle of $60^{\circ}$). While data were recorded over extended regions in momentum space, here we focus on the trARPES signal in the vicinity of the high symmetry point $\bar{K}$, a corner of the hexagonal surface-projected Brillouin zone (fig. \ref{fig:PESsummary}d). For the experiment, a split portion of the fundamental output of the OPCPA at 1.55~eV was used as pump, with a peak fluence on the sample of 1.1~mJ/cm$^2$. The trARPES spectrum is shown in figure \ref{fig:PESsummary}a, before (left) and during (right) pump-probe temporal overlap. The energy zero is set for convenience to the local maximum of the valence band at $\bar{K}$, an inherently two-dimensional state\cite{bertoni2016}. 
During temporal overlap between the XUV pulse and the 1.55~eV s-polarized pump, non-resonant two-photon photoemission from the spin-orbit split valence band produces a short-lived signal in the material's band gap, separated by 1.55~eV from the initial valence band states\cite{hein2016}. We use this non-resonant signal to estimate the temporal cross-correlation between pump and probe: the intensity in a small region at 1.1~eV and -1.27~\AA$^{-1}$ (blue box in figure \ref{fig:PESsummary}a) is shown in figure \ref{fig:PESsummary}b. The maximum of this signal is used as a zero for the temporal axis and the trace is normalized. This signal is fitted with a Gaussian function: the curve's FWHM is $37\pm2$~fs. This value can be compared with the pump pulse duration of 32~fs, which was characterized by frequency-resolved optical grating: the system response function is apparently dominated by the pump pulse duration. Assuming both pump and probe pulses to have a Gaussian envelope, with a time-bandwidth product of $1824$ meV$\cdot$fs, one obtains a probe pulse width of 19~fs FWHM, not far from the Fourier-limited pulse duration of 17~fs extracted from the spectral width of 110~meV. This suggests that an even better temporal resolution, without losses in energy resolution, would be possible by further shortening of the pump pulses. The dataset, acquired with an integration time of 31~s per delay point, clearly demonstrates the ability of the setup to follow population dynamics of excited states, unoccupied at equilibrium, in the entire Brillouin zone. The detectable in-gap two-photon photoemission signal has an intensity of $\approx10^{-4}$ relative to the simultaneously-measured VB: this high dynamic range can be used to follow in a single data acquisition the evolution of states below and above the Fermi level. The excited states in this proof-of-principle experiment show rich momentum dependent relaxation dynamics which will be discussed in detail in a future publication. 

The experimental energy resolution was checked on a sample with bands crossing the Fermi level, TbTe$_{3}$: a momentum-integrated energy distribution curve (EDC) is shown in figure \ref{fig:PESsummary}c; the data set was collected at a temperature of 100 K, with a photocurrent of 130 pA, corresponding to 1.6$\times10^{3}$ electrons emitted per pulse. The EDC was fitted by a Fermi-Dirac distribution function with the temperature of 100 K as fixed parameter, convolved by a Gaussian broadening as fit parameter: the total resolution is 121 meV. This value includes the dominant 110 meV source's linewidth, broadened by the analyzer resolution of 35 meV and a residual space-charge contribution, on the order of 35-40 meV. 

\section{\label{sec:conclusion} Conclusions}
In conclusion we have demonstrated a novel setup for high-repetition-rate trARPES at XUV energies, capable of band mapping in the whole Brillouin zone of most materials. The light source developed for the experiment is based on a Ytterbium-based OPCPA operating at 500~kHz, frequency up-converted to the 21.7~eV by a UV-driven HHG source. Time-resolved two-photon photoemission studies are feasible in a broad reciprocal space region with a system response function below 40~fs. The novel setup can map the excited-state band structure and follow its evolution on a femtosecond time-scale\cite{Nicholson2018}.

\section*{Acknowledgements}
We thank S.~Kubala, M.~Krenz, D.~Bauer, R.~Franke, P.~Heyne, W.~Erlebach., J.~Malter and C.~Cacho for technical support. This work was funded by the Max Planck Society, the European Research Council (ERC) under the European Union{'}s Horizon 2020 research and innovation program (Grant No.~ERC-2015-CoG-682843), the European Union{'}s FP7 program through the CRONOS project, Grant No.~280879, and the German Research Foundation (DFG) within the Emmy Noether program (Grant No. RE 3977/1). C.~M.~acknowledges support by the Swiss National Science Foundation under Grant No.~PZ00P2-154867.

\bibliography{REVSCINT_FHIHHG}

\end{document}